\documentstyle[aps,prb,psfig,preprint]{revtex}
\draft
\begin{document}

\title{Characteristic features of d pairing in bilayer cuprates
under conditions of Peierls instability of the normal phase}
\author{M. V. Eremin$^{*}$ and I. A. Larionov}
\address{ Physics Department, Kazan State University, 420008 Kazan, Russia}
\maketitle


\begin{abstract}

A system of self-consistent integral equations for the superconducting
gap is formulated and solved taking account of the instability of the
normal phase of bilayered cuprates against charge-density waves. The
critical parameters are calculated as a function of the wave vector,
temperature, and doping index. It is found that the region in which
superconductivity coexists with d-type charge-density waves depends
strongly on the doping index. The effective energy-gap parameter,
determined as the interval between the peaks of the density of states, can
have a local minimum at temperatures T$<$T$_C$.
\newline

\end{abstract}

Journal ref:

Pis'ma Zh. Eksp. Teor. Fiz. 68, No. 7, 583-587, (1998)

JETP Lett. vol. 68, \# 7, pp. 611-615,  10 October 1998
 \\
 \\
$^*$ corresponding author, e-mail: Mikhail.Eremin@ksu.ru
\pacs{PACS numbers: 74.72.-h, 74.25.Jb}

It can now be regarded as an established experimental fact that
d-type pairing is realized in most layered cuprates, and in addition
the determining component of the order parameter of the pseudogap in
the normal phase of underdoped samples possesses the same d-type
symmetry.$^{1,2}$ At least this conclusion does not raise any objections
for YBa$_2$Cu$_3$O$_{7-\delta}$, Bi$_2$Sr$_2$CaCu$_2$O$_{8-\delta}$,
and other bilayer cuprates. The question of the nature of the pairing
and the origin of the pseudogap in the normal phase  is still open.
Specifically, it is still not clear whether or not they are interrelated
or completely independent. In the present letter we show that the
empirical dependences of $T_C$ and the pseudogap closure temperature
$T^*$ on the doping index and the dependences  of their
order parameters on the wave vector can be qualitatively understood in
a model of a quasi-two-dimensional metal with strong electronic
correlations, in which case the gap in the elementary excitations
spectrum is due to anomalous averages of the type
$\langle \Psi_{\bf k}^{2 \sigma} \Psi_{\bf k+q}^{2 \sigma} \rangle$,
characteristic for a Peierls instability of the normal phase.
The problem of the existence of superconductivity under conditions
of Peierls instability is a classic problem. For ordinary metals
it has been discussed in a number of works (see, for example,
Refs. 3 and 4), where the anomalous averages were due to the
Fr\"ohlich interaction and did not depend on the wave vector.
In the present letter the main interaction responsible for the
formation of the superconducting state is the short-range
superexchange interaction, and the normal phase exhibits
strongly non-Fermi-liquid behavior.
The spectral weight of the conduction band under study depends
nonlinearly on the doping index, and it is already half-filled
when the number of holes per unit cell of the
Cu$_2$O$_4$ bilayer is 2/7 (Ref. 5) instead of 1,
as would have happened in the case of normal metals.

To describe the subsystem of current carriers in bilayer
high-$T_C$ superconductors we employ a Hamiltonian of the form
\begin{equation}
\hat{H}= \sum_{l\sigma } \varepsilon _p\Psi _l^{\sigma,\sigma }+
\sum_{lm}t_{lm}^{pp}\Psi _l^{2,\sigma }\Psi_m^{\sigma,2}
+ \sum_{l>m}j_{lm}\left[ 2\left( \overline{S}_l\overline{S}_m\right)
-\frac{n_ln_m}2 \right]  + \sum_{l>m}g_{lm}\delta_l\delta _m,
\label{Ham}
\end{equation}
where $\Psi_m^{2 \sigma} (\Psi_m^{\sigma 2} )$,
are Hubbard-type operators corresponding to creation (annihilation) of
quasiparticles in the bonding singlet-correlated oxygen band,$^6$
$n_l = \Psi_l^{ \uparrow \uparrow} + \Psi_l^{\downarrow \downarrow} $,
$\delta_l = n_l+2\Psi_l^{2 2} $,  and, $j_{lm}$ and $g_{lm}$ are
the superexchange and Coulomb interactions parameters.
The quasiparticle dispersion relation has the form
\begin{equation}
\varepsilon_{\bf k} = P_{pd} [2t_1 (\cos {\bf k}_x a + \cos {\bf k}_y a) +
4t_2 \cos {\bf k}_x a \cos {\bf k}_y a +
2t_3 (\cos 2{\bf k}_x a + \cos 2{\bf k}_y a)]- \mu,
\end{equation}
where $\mu$ is the chemical potential, which under optimal doping,
according to photoemission
data,$^7$ lies 10 meV below the saddle peak of the density of states,
$P_{pd} = (2+\delta)/4 $ is the average value of the anticommutators
 $\lbrace \Psi_m^{2 \sigma}, \Psi_m^{\sigma 2} \rbrace$
 with tunneling taken into account,
$\delta$ is the number of current carriers per unit cell in the
Cu$_2$O$_4$ bilayer, $a$ is the lattice constant, and
$t_1 =t_1^{(0)} [1+\langle S_i S_j \rangle /P_{pd}^2 $.
 We take the following parameter set, which is
consistent with the Fermi surface and the temperature behavior
of the spin susceptibility of the normal phase of YBa$_2$Cu$_4$O$_8$:
$ t_1^{(0)}$ = 70 meV, $t_2^{(0)}$ = 0, and $t_3^{(0)}$ = 5 meV.$^8$
 Following Ref. 9, we give the spin correlation functions of the nearest
neighbors phenomenologically
as $\langle S_i S_j \rangle = (-0.2 \exp(0.6+ \delta/2)^{-1} +0.6)/1.8356$,
so that the width  of the band would approach zero as
$\delta \rightarrow 0$, which is necessary for a correct description of the
insulator-metal transition in these compounds.

In the mean-field approximation the elementary excitations
spectrum of the Hamiltonian (1) is determined by the equation

\begin{equation}
\det \left|
\begin{tabular}{llll}
\mbox{\hspace{5mm}}$ \varepsilon_{\bf k}-E$ & \mbox{\hspace{9mm}}$G_{\bf k+Q}$ &
\mbox{\hspace{12mm}}$ \Delta_{\bf k}$ & \mbox{\hspace{12mm}}$U_{\bf k}$ \\
\mbox{\hspace{9mm}}$ G_{\bf k}$ & \mbox{\hspace{5mm}}$\varepsilon _{\bf k+Q}-E$ &
\mbox{\hspace{9mm}}$ U_{\bf k+Q}$ & \mbox{\hspace{9mm}}$\Delta_{\bf k+Q}$ \\
\mbox{\hspace{9mm}}$ \Delta_{\bf k}^{*}$ & \mbox{\hspace{9mm}}$U_{\bf k+Q}^{*}$ &
\mbox{\hspace{5mm}}$ -\varepsilon_{\bf k}-E$ & \mbox{\hspace{9mm}}$-G_{\bf k+Q}^{*}$ \\
\mbox{\hspace{9mm}}$ U_{\bf k}^{*}$ & \mbox{\hspace{9mm}}$\Delta_{\bf k+Q}^{*}$ &
\mbox{\hspace{9mm}}$ -G_{\bf k}^{*}$ & \mbox{\hspace{5mm}}$-\varepsilon_{\bf k+Q}-E$\mbox{\hspace{5mm}}
\end{tabular}
\right| =0.
\end{equation}

The order parameter of the superconducting transition
(ST) is given by the expression
\begin{equation}
\Delta_{{\bf k}_1} = \frac {1}{P_{pd}N}\sum_{{{\bf k}}}
[2j({\bf k}_1-{\bf k}) - g({\bf k}_1-{\bf k}) +
 P^2_{pd} B({{\bf k}}_1,{{\bf k}}_1-{{\bf k}}) ]
\langle \Psi _{{{\bf k}}}^{\downarrow, 2}
\Psi_{-{{\bf k}}}^{\uparrow,2} \rangle ,
\label{SCdef}
\end{equation}
where $j({\bf q})$, $g({\bf q})$, and $B({\bf k}_1,{\bf q})$
are the Fourier transforms of the superexchange, Coulomb,
and Fr\"ohlich interactions potentials, respectively. Specifically,
$$j({\bf q})=j_0 (1-6 \delta^2 )\cdot(\cos q_x a + \cos q_y a),$$
where $j_0$ = 125 meV for YBaCuO compounds$^{10}$ and
$$g({\bf q})=g_0 \exp(-7 \delta)(\cos q_x a + \cos q_y a),$$
where $g_0$ = 315 meV. Among all possible phonon
modes, the buckling modes make the most important contribution
to pairing. In this case

\begin{equation}
B({\bf k},{\bf q}) = 2B_0^2 \hbar \omega_{\bf q}
\frac {1+\frac 12 [ \cos q_x a + \cos q_y a]}
{(\hbar \omega_{\bf q})^2 -(\varepsilon_{\bf k} -\varepsilon_{\bf k+q} )^2}
\Theta(\hbar \omega_D-\left|\varepsilon_{\bf k}-\varepsilon_{\bf k+q}\right|)
\Theta(\hbar \omega_D-\left|\varepsilon_{\bf k}\right|),
\label{phonon}
\end{equation}
$\hbar \omega_D$ is the Debye frequency of the order of 45 meV,
$\Theta(x)$ is the theta function, and $B_0 \simeq$ 35 meV.
In the range of doping indices $\delta$ of interest to us,
 Eq. (4) has a solution of the form
$\Delta_{\bf k} = \Delta_0(\cos k_x a - \cos k_y a)$.

As was pointed out in Ref. 5, the CDW parameter $G_{\bf k}$ is a sum
of two components with a different dependence on the wave vector.
The component associated with the short-range potentials

\begin{equation}
G_{{\bf k}_1}^{ex} = \frac {1}{P_{pd}N} \sum_{\bf k}
\lbrack j({\bf k}_1-{\bf k}) + g({\bf k}_1-{\bf k}) \rbrack
\langle \Psi_{{{\bf k}}}^{2,\uparrow}
\Psi _{{{\bf k}}+{{\bf Q}}}^{ \uparrow,2} \rangle,
\label{Gexdef}
\end{equation}
possesses d-type symmetry $G_{\bf k}^{ex}=iG_0 (\cos k_x a -\cos k_y a)$,
 while the phonon part
\begin{equation}
G_{{\bf k}_1}^{ph} = \frac {P_{pd}}{N} \sum_{\bf k}
V_{{\bf k},{\bf Q}} \langle \Psi_{{\bf k}+{\bf Q}}^{2 \uparrow}
\Psi_{\bf k}^{\uparrow 2} \rangle +
V_{{\bf k}_1,{\bf Q}} \frac {P_{pd}}{N} \sum_{\bf k}
\langle \Psi_{{\bf k}+{\bf Q}}^{2 \uparrow}
\Psi _{\bf k}^{ \uparrow 2} \rangle,
\label{Gph}
\end{equation}
possesses s-type symmetry. It is relatively small and is neglected below.
The order parameter $U_k$ entering in Eq. (3) is related with
thermodynamic average $\langle \Psi_{{{\bf k}}}^{\downarrow 2}
\Psi_{{-{\bf k}}-{{\bf Q}}}^{\uparrow 2}\rangle $, which appears naturally
 in the equations of motion in the presence of
superconductivity and Peierls instability.
In the general case it is determined by the expression
\begin{equation}
U_{{\bf k}_1} = \frac{1}{P_{pd}N} \sum_{{\bf k}}
[j({\bf k}_1 - {\bf k}) + j({\bf k}_1 - {\bf k} - {\bf Q})
- g({\bf k}_1 - {\bf k}) + P_{pd}^2 B({\bf k_1},{\bf k}_1 - {\bf k}) ]
\langle \Psi_{{{\bf k}}}^{\downarrow 2}
\Psi_{{-{\bf k}}-{{\bf Q}}}^{\uparrow 2} \rangle.
\end{equation}
We investigate below the approximately optimal doping regime,
where $Q \simeq (\pi,\pi)$. One can see from Eq. (8) that $U_k$
is comparatively small and can possess only s symmetry,
since for $Q=(\pi,\pi)$ the sum
$j({\bf k}_1 -{\bf k})+j({\bf k}_1-{\bf k}-{\bf Q})$ vanishes.
We note that for $U_{\bf k}$ = 0 Eq. (3) leads to the spectrum
\begin{equation}
E_{1,2}=\sqrt{\frac 12 (\varepsilon^2_{\bf k} + \varepsilon^2_{\bf k+Q})
+\Delta_{\bf k}\Delta^*_{\bf k}
+ G_{\bf k} G^*_{\bf k} \pm \frac 12 E^2_{12} }
\end{equation}
where
\begin{equation}
E^2_{12}=\sqrt{(\varepsilon^2_{\bf k} - \varepsilon^2_{\bf k+Q} )^2
 + 4  G_{\bf k} G^*_{\bf k} (\varepsilon_{\bf k}+\varepsilon_{\bf k+Q})^2
+ 4\Delta_{\bf k}\Delta^*_{\bf k} (G_{\bf k} + G^*_{\bf k})^2}
\end{equation}
and $E_4=-E_1$ and $E_3=-E_2$.

In the general case we have the following system of integral equations:
\begin{eqnarray}
\langle \Psi_{{{\bf k}}}^{\downarrow 2}
\Psi_{{-{\bf k}}}^{\uparrow 2} \rangle =\frac{P_{pd}\Delta_{\bf k}}{4}
\left[ \frac{1}{E_1}\tanh \left(\frac{E_1}{2k_B T} \right) +
\frac{1}{E_1}\tanh \left(\frac{E_1}{2k_B T} \right)  \right]  \nonumber \\
+ \frac{P_{pd}N_{\Delta}}{2E^2_{12}}
\left[ \frac{1}{E_1}\tanh \left(\frac{E_1}{2k_B T} \right) -
\frac{1}{E_1}\tanh \left(\frac{E_1}{2k_B T} \right)  \right], \nonumber \\
\langle \Psi_{{{\bf k}}}^{2 \uparrow }
\Psi_{{{\bf k}+{\bf Q}}}^{\uparrow 2} \rangle =\frac{P_{pd}G_{\bf k}}{4}
\left[ \frac{1}{E_1}\tanh \left(\frac{E_1}{2k_B T} \right) +
\frac{1}{E_1}\tanh \left(\frac{E_1}{2k_B T} \right)  \right]  \nonumber \\
+ \frac{P_{pd}N_{G}}{2E^2_{12}}
\left[ \frac{1}{E_1}\tanh \left(\frac{E_1}{2k_B T} \right) -
\frac{1}{E_1}\tanh \left(\frac{E_1}{2k_B T} \right)  \right], \nonumber \\
\langle \Psi_{{{\bf k}}}^{\downarrow 2}
\Psi_{{-{\bf k}-{\bf Q}}}^{\uparrow 2} \rangle =\frac{P_{pd}U_{\bf k}}{4}
\left[ \frac{1}{E_1}\tanh \left(\frac{E_1}{2k_B T} \right) +
\frac{1}{E_1}\tanh \left(\frac{E_1}{2k_B T} \right)  \right]  \nonumber \\
+ \frac{P_{pd}N_{U}}{2E^2_{12}}
\left[ \frac{1}{E_1}\tanh \left(\frac{E_1}{2k_B T} \right) -
\frac{1}{E_1}\tanh \left(\frac{E_1}{2k_B T} \right)  \right],
\end{eqnarray}
where
\begin{eqnarray}
N_{\Delta}=\frac 12 \Delta (\varepsilon^2_{\bf k} - \varepsilon^2_{\bf k+Q})
+ \Delta_{\bf k}G^*_{\bf k}(G_{\bf k}+G^*_{\bf k})+U_{\bf k}
(\Delta_{\bf k}U^*_{\bf k}-\Delta^*_{\bf k}U_{\bf k})
+ 2\varepsilon_{\bf k+Q} G^*_{\bf k}U_{\bf k}, \nonumber \\
N_G=\frac 12 G_{\bf k}(\varepsilon_{\bf k} + \varepsilon_{\bf k+Q})^2
+ \Delta_{\bf k}\Delta^*_{\bf k}(G_{\bf k}+G^*_{\bf k})
+ \varepsilon_{\bf k+Q}\Delta^*_{\bf k}U_{\bf k}
- \varepsilon_{\bf k}\Delta_{\bf k}U^*_{\bf k}, \nonumber \\
N_U=\frac 12 U_{\bf k}(\varepsilon_{\bf k} - \varepsilon_{\bf k+Q})^2
-\Delta_{\bf k}(\Delta_{\bf k}U^*_{\bf k}-\Delta^*_{\bf k}U_{\bf k})
- \varepsilon_{\bf k}\Delta_{\bf k}G^*_{\bf k}
+ \varepsilon_{\bf k+Q}\Delta_{\bf k}G_{\bf k}.
\end{eqnarray}

The system of equations (11)-(13) was solved numerically. As expected,
the parameter $U_k$ is negligibly small. The computed phase diagram of
the $\delta$ dependence of the critical temperatures for $\Delta_0 (T_C)$
and $G_0 (T^*)$ at $Q=(\frac{11}{12} \pi, \frac{11}{12} \pi)$ is close
to the corresponding diagrams proposed in experimental works (see,
for example, Ref. 11). In Fig. 1 three different doping cases are
presented as an example of the temperature dependences found for
the amplitudes of the order parameters $\Delta_0 (T)$ and $G_0 (T)$:
$\alpha_1$, $\alpha_2$ and $\alpha_3$, where the parameter $\alpha$
is given by the ratio
$\alpha(\delta)=(\delta - \delta_{min})/(\delta_{opt} - \delta_{min})$.
Here $\delta_{min}$ is the minimum value of the doping corresponding
to the appearance of a superconducting state. In all cases, for
$\delta < \delta_{opt}$ = 0.315 (i.e., the existence of a CDW)
the temperature dependence of the superconducting order parameter
differs considerably from the BCS case. In the case of a strongly
overdoped regime the ratio $4\Delta_0 / k_B T_C$ = 4.2; for $\delta_{opt}$
it equals 4.5; it then increases, as seen in Fig. 1, and reaches values
of the order of 10 for $\delta \gtrsim \delta_{min}$ (i.e., in the
strongly underdoped region). This kind of variation of $4\Delta_0/k_B T_C$
has already been noted in a number of experimental works.$^{12}$
The computed temperature dependences of the pseudogap $G_0$ exhibit
nonmonotonic behavior as temperature decreases; this has not been
noticed previously. It would be interesting to check this prediction
of the theory experimentally. Thus, if the effective gap parameter
$\Delta_{eff} (T)$, determined as the energy interval between peaks
in the density of states, is measured in scanning tunneling
spectroscopy experiments (see, for example, Ref. 12), then on account
of the nonmonotonic behavior of $G_0 (T)$ anomalies should be observed
in the plot for $\Delta_{eff} (T)$: 1) In type $\alpha_1$ samples
$\Delta_{eff} (T)$ will decrease (!) as $T \rightarrow 0$;
2) for $\alpha \approx \alpha_3$ a local
minimum should be observed in the plot of $\Delta_{eff} (T)$.
Examples of such plots calculated on the basis of expressions (9) and (10)
are presented in Fig. 2. In our opinion the indicated
anomalies can be seen from the experimental data.$^{12,13}$
These results, however, require a more detailed experimental
check on samples with different doping indices.

This work was supported in part by the Russian State Science and Technology
Program "Superconductivity", Project No. 98014.

\begin{figure}[t]
%
%
\centerline{\psfig{clip=,file=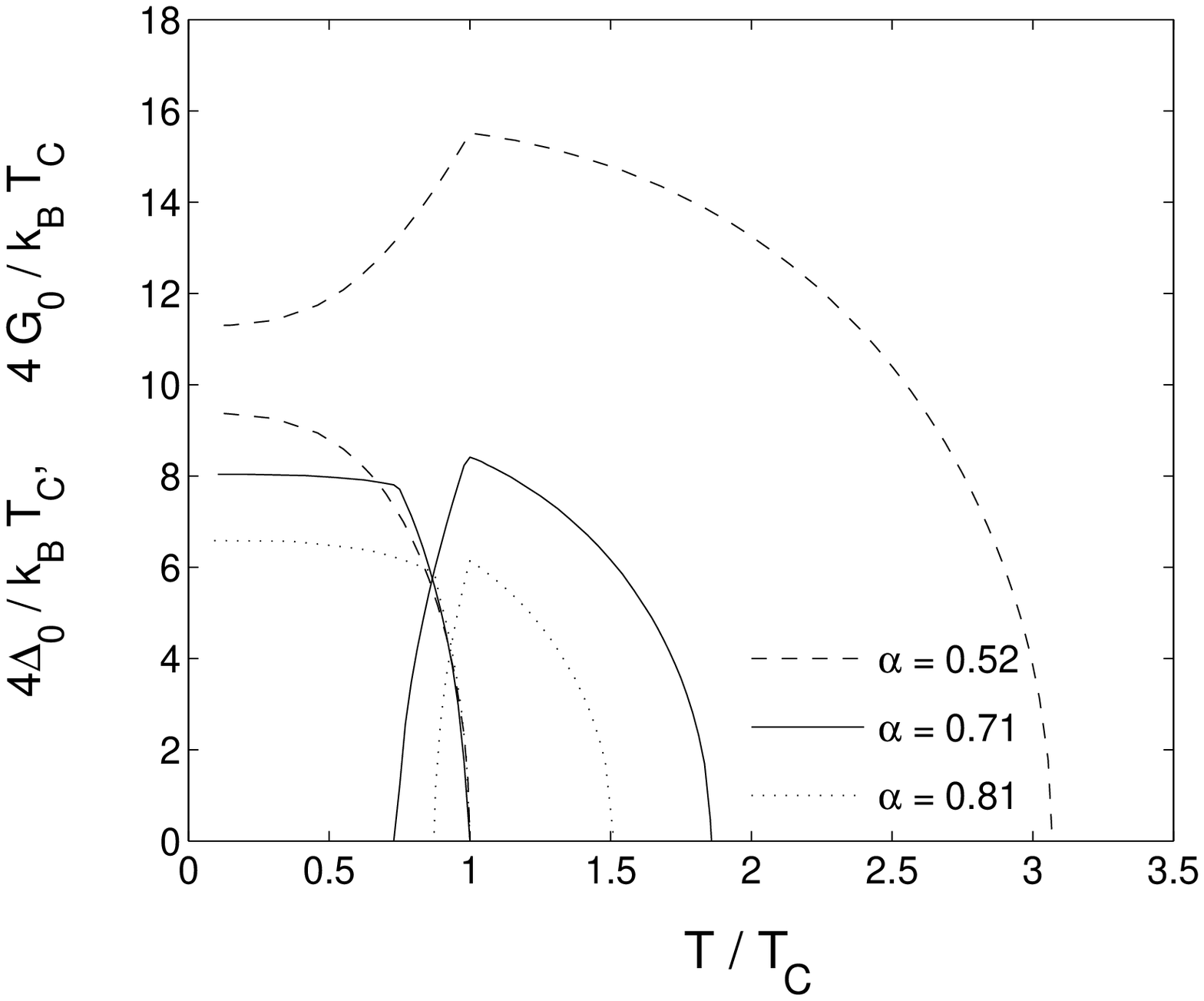,width=10.7cm,angle=0}}
%
\caption{Computed temperature dependences of the amplitudes of the order
parameters $\Delta_0$ and $G_0$. The curves corresponding to the
superconducting gap parameter $\Delta_0$ start at the point $T/T_C = 1$.
The relative-doping parameter $\alpha$ is defined in the text.
}
\label{fig1}
\end{figure}
\begin{figure}[t]
%
%
\centerline{\psfig{clip=,file=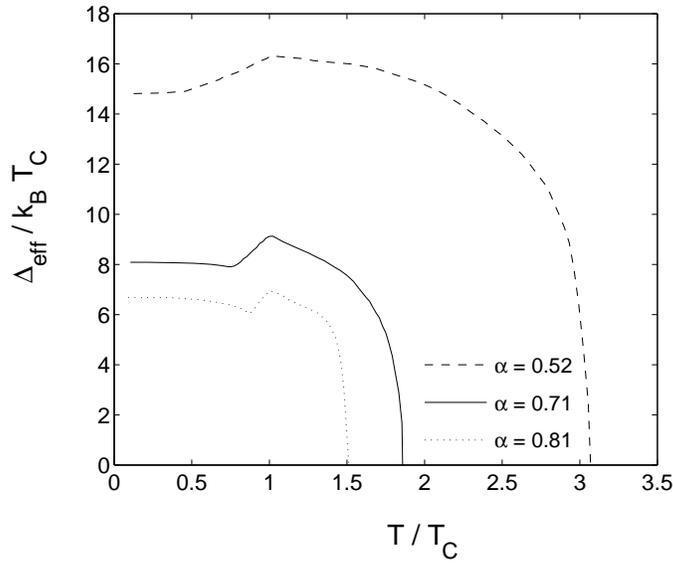,width=10.7cm,angle=0}}
%
\caption{Predicted examples of nonmonotonic temperature variations
of the effective gap $\Delta_{eff}$ between peaks of
the density of states of the energy spectrum.
}
\label{fig2}
\end{figure}

\end{document}